\documentclass[twocolumn,superscriptaddress,english,pre,showpacs,longbibliography]{revtex4-2}

\usepackage{graphicx}
\usepackage{dcolumn}
\usepackage{bm}
\usepackage{subfigure}
\usepackage{hyperref}
\usepackage{url}
\usepackage{amsthm,amsmath,amssymb}
\usepackage{mathrsfs}
\usepackage{color}

\begin{document}
\preprint{APS/123-QED}

\title{A method for quantifying the generalization capabilities of generative models\\ for solving Ising models}

\author{Qunlong Ma}
\affiliation{Henan Key Laboratory of Network Cryptography Technology, Zhengzhou 450001, China}

\author{Zhi Ma}
\affiliation{Henan Key Laboratory of Network Cryptography Technology, Zhengzhou 450001, China}

\author{Ming Gao}
\email{gaoming@nudt.edu.cn}
\affiliation{Henan Key Laboratory of Network Cryptography Technology, Zhengzhou 450001, China}

\date{\today}

\begin{abstract}
For Ising models with complex energy landscapes, whether the ground state can be found by neural networks depends heavily on the Hamming distance between the training datasets and the ground state. Despite the fact that various recently proposed generative models have shown good performance in solving Ising models, there is no adequate discussion on how to quantify their generalization capabilities. Here we design a Hamming distance regularizer in the framework of a class of generative models, variational autoregressive networks (VAN), to quantify the generalization capabilities of various network architectures combined with VAN. The regularizer can control the size of the overlaps between the ground state and the training datasets generated by networks, which, together with the success rates of finding the ground state, form a quantitative metric to quantify their generalization capabilities. We conduct numerical experiments on several prototypical network architectures combined with VAN, including feed-forward neural networks, recurrent neural networks, and graph neural networks, to quantify their generalization capabilities when solving Ising models. Moreover, considering the fact that the quantification of the generalization capabilities of networks on small-scale problems can be used to predict their relative performance on large-scale problems, our method is of great significance for assisting in the Neural Architecture Search field of searching for the optimal network architectures when solving large-scale Ising models.

\end{abstract}
\maketitle

\section{INTRODUCTION}

When solving intractable and NP-hard Ising models \cite{Barahona1982} using classical computers, heuristic algorithms \cite{Kirkpatrick1983, PT1986} are a common alternative to exhaustive search, which tends to grow exponentially with system size. For Ising models with complex energy landscapes, the probability that heuristic algorithms fall into one of the many excited states with energy very close to the energy of the ground state is particularly high. Only when there is a large overlap between the initial state and the ground state can the algorithms obtain the ground state. Therefore, whether the ground state can be found by heuristic algorithms depends heavily on the Hamming distance between the initial state and the ground state \cite{Perdomo2011, Albash2021, Valle2021, Mehta2022}.

Many deep neural networks have been proposed to solve Ising models recently \cite{Carleo2019, Akinori2023}, one class of which is generative models \cite{Wu2019VAN, Hibat2021vca, McNaughton2020, Marylou2022, Wu2021, Panfeng2021, Pixel2016,qunlong2023}. Corresponding to classical heuristic algorithms, whether the ground state can be found by generative models depends heavily on the Hamming distance between the training datasets and the ground state. In deep learning literature, the generalization of neural networks generally refers to classifying or generating data that have never been seen before and satisfy certain conditions during training \cite{Goodfellow2016}.

Although many generative models perform well in solving Ising models, there is no adequate discussion on how to quantify their generalization capabilities. In the deep learning and quantum machine learning communities, there are some discussions of the generalization capabilities of networks when used for classification tasks and graph generation tasks \cite{huang2022towards, Zhou2022, lee17a, mallasto2020, thompson2022, Banchi2021, du2022power, gili2023quantum}. To the best of our knowledge, there is no similar research or discussion when networks are used for solving intractable Ising models. It is harder to find suitable quantitative metrics and canonical datasets to compare the generalization capabilities of network architectures when used for solving Ising models compared with classification tasks or graph generation tasks.

\begin{figure*}
\includegraphics[scale=0.265]{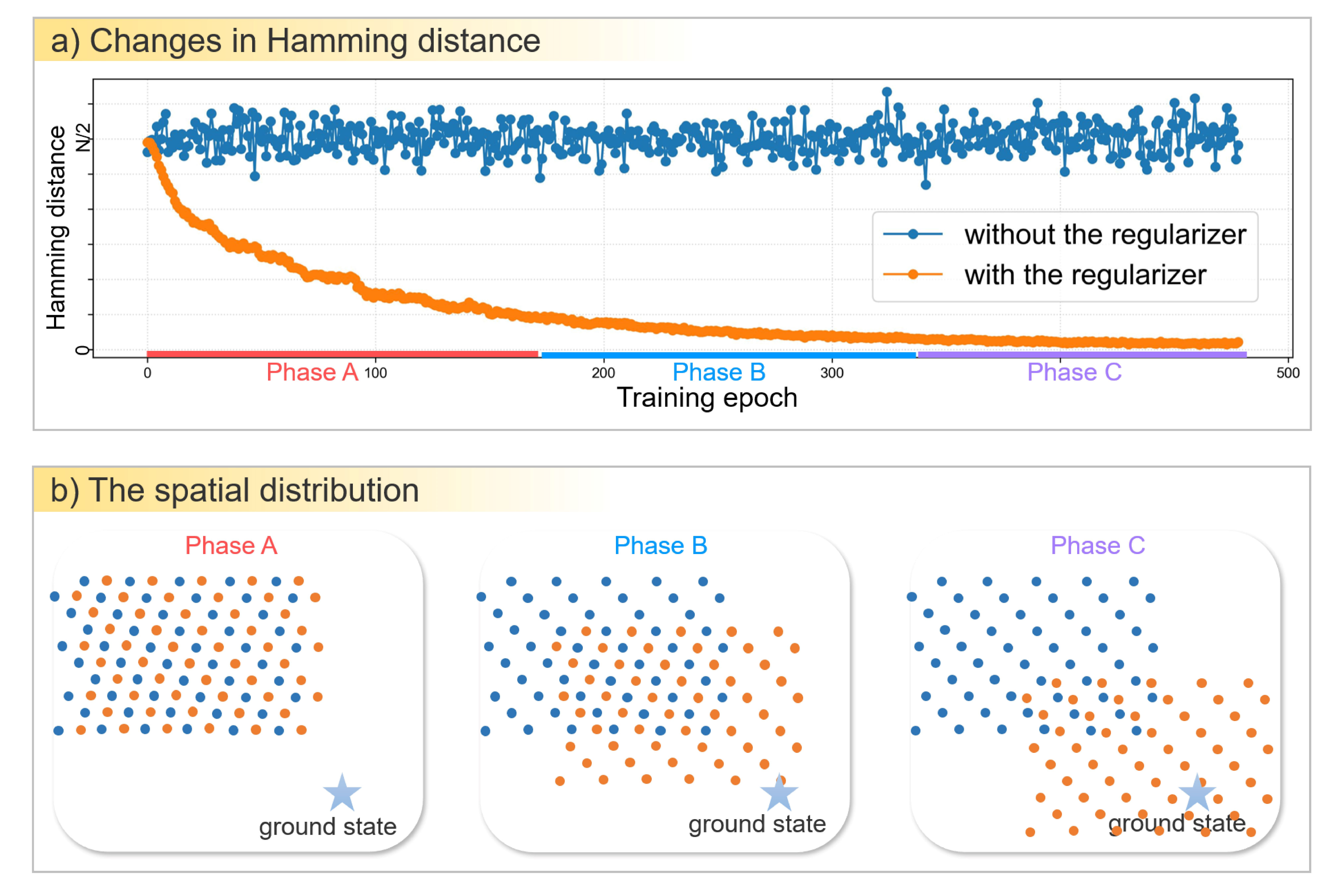}
\caption{\label{fig1} Schematic diagram of Hamming distance and configuration distribution trained with and without the regularizer in the loss function. (a) The changes in average Hamming distance between the ground state and the training datasets, which are directly sampled from networks during training, when with and without the regularizer in the loss function. (b) The spatial distribution of configurations sampled directly from networks at three phases during training, when with and without the regularizer in the loss function.}
\end{figure*}

In this work, we design a Hamming distance regularizer in the framework of a class of generative models, variational autoregressive networks (VAN) \cite{Wu2019VAN} to quantify the generalization capabilities of various network architectures combined with VAN. Since the training datasets are generated by networks, we can impose restrictions on networks to generate training datasets that satisfy certain special properties. Thus, we can flexibly control the size of the overlaps between the ground state and the training datasets by adjusting the Hamming distance regularizer and compare the success rates of finding the ground state among network architectures. We then propose a quantitative metric that combines the size of the overlaps with the success rates to quantify the generalization capabilities of network architectures. Our method can be used to assist in the Neural Architecture Search (NAS) field \cite{Elsken2019, wistuba2019, mellor21a}, which is to automatically search for the optimal network architectures for solving a class of Ising models instead of hand-designing them.

We conduct numerical experiments on several prototypical network architectures combined with VAN, such as feed-forward neural networks (FNN) \cite{Goodfellow2016}, recurrent neural networks (RNN) \cite{Hibat2020RNN}, and graph neural networks (GNN) \cite{semi2017GCN3}, to quantify the generalization capabilities when solving disordered, fully connected Ising models, including the Wishart planted ensemble (WPE) \cite{WPE2020} and the Sherrington-Kirkpatrick (SK) model \cite{Sherrington1975SK}. We also consider the effect of a hyperparameter, the number of layers in network architectures, on generalization capabilities. In addition, considering the fact that the quantification of the generalization capabilities of networks on small-scale problems can be used to predict their relative performance on large-scale problems, our method is of great significance for assisting in the NAS field of searching for the optimal network architectures when solving large-scale Ising models.

The paper is structured as follows. In Sec.~\ref{sec2}, we describe in detail the Hamming distance regularizer and how it can be used to quantify the generalization capabilities of network architectures. In Sec.~\ref{sec3}, we conduct numerical experiments to quantify the generalization capabilities of network architectures combined with VAN on the WPE and the SK model at multiple system sizes. We conclude and discuss in Sec.~\ref{sec4}.

\section{THE HAMMING DISTANCE REGULARIZER}
\label{sec2}

The loss function of our method is composed of the variational free energy $F_q$ and the Hamming distance regularizer $R_h$,
\begin{equation}
    \label{eq1}
    \mathcal{L} = F_q + R_h.
\end{equation}

The variational free energy $F_q$ is the same as defined in VAN \cite{Wu2019VAN}, which is
\begin{equation}
    \label{eq2}
    F_q=\sum_{\textbf{s}}{q_{\theta}(\textbf{s})\left[{E}(\textbf{s})+\frac{1}{\beta}\ln{q_{\theta}(\textbf{s})}\right]},
\end{equation}
where $\beta$ is inverse temperature, and ${E}(\textbf{s})$ denotes the Hamiltonian of the configuration $\textbf{s}$. The variational distribution $q_{\theta}(\textbf{s})$ is parameterized by $\theta$ of networks. In existing methods based on VAN \cite{Wu2019VAN, Hibat2021vca, Panfeng2021, qunlong2023}, $F_q$ is directly used as the loss function. More details about the $F_q$ can be found in Ref.~\cite{Wu2019VAN, Hibat2021vca, qunlong2023}.

The Hamming distance regularizer is written as
\begin{equation}
    \label{eq3}
    R_h=\sum_{\textbf{s}}{|hm_{\textbf{g}}(\textbf{s})-z|},
\end{equation}
where $hm_{\textbf{g}}(\textbf{s})$ denotes the Hamming distance between the ground state $\textbf{g}$ and the configuration $\textbf{s}$, and $z$ is a constant to reflect how big the overlap of them is, that is, the \textit{spatial distance}.

After calculating the variational free energy and the regularizer, we can get the gradient of the loss function about the parameter ${\theta}$,
\begin{equation}
\begin{aligned}
    \label{eq4}
\bigtriangledown_{\theta}\mathcal{L}={\mathbb{E}_{\textbf{s}\sim q_{\theta}(\textbf{s})}\left\{\left[{E}(\textbf{s})+\frac{1}{\beta}\ln{q_{\theta}(\textbf{s})}\right]\bigtriangledown_{\theta}\ln{q_{\theta}(\textbf{s})}\right\}},
\end{aligned}
\end{equation}
which is independent of the Hamming distance regularizer, as it is in Ref.~\cite{Wu2019VAN,Hibat2021vca,qunlong2023}.

Since all the methods we compare in this paper are based on the VAN framework, which are trained in the same way \cite{Wu2019VAN,Hibat2021vca,qunlong2023}, i.e. the only differences between them are network architectures. More details about how to train them can be found in Ref.~\cite{Wu2019VAN,Hibat2021vca,qunlong2023}.

The previous researches have illustrated that only by containing the configurations in the training datasets that are close to the ground state, measured by Hamming distance, to train the neural networks, may we obtain the ground state after training \cite{Perdomo2011, Albash2021, Valle2021, Mehta2022}. Therefore, we design this regularizer to explore the relationship between the Hamming distance and the success rates of finding the ground state for different network architectures combined with VAN. Also, it is not only related to the different network architectures and training hyperparameters but also to the characteristics of benchmark problems.

In Fig.~\ref{fig1}(a), we show a schematic diagram of the changes in the average Hamming distance between the ground state and the training datasets, which is directly sampled from networks during training, when with or without the regularizer in the loss function. In Fig.~\ref{fig1}(b), we demonstrate a schematic of the spatial distribution of configurations directly sampled from networks during three phases in Fig.~\ref{fig1}(a). As training proceeds, the spatial distribution of configurations sampled from networks trained with the regularizer in the loss function tends towards the ground state, and the configurations sampled from networks trained without the regularizer are always heterogeneous.

\section{NUMERICAL EXPERIMENTS}
\label{sec3}

Our method is achieved by controlling the size of the overlaps between the ground state and the training datasets sampled directly from networks. Thus, we can use the size of the overlaps and the corresponding success rates of finding the ground state to quantify the generalization capabilities of network architectures. In this work, we focus on the generalization capabilities of the VAN framework combined with various network architectures, the effect of the number of layers in networks, and whether our method performs consistently on both large-scale and small-scale problems, that is, whether we can predict the relative performance of network architectures on large-scale problems using their performance on small-scale problems.

\begin{figure*}
\includegraphics[scale=0.38]{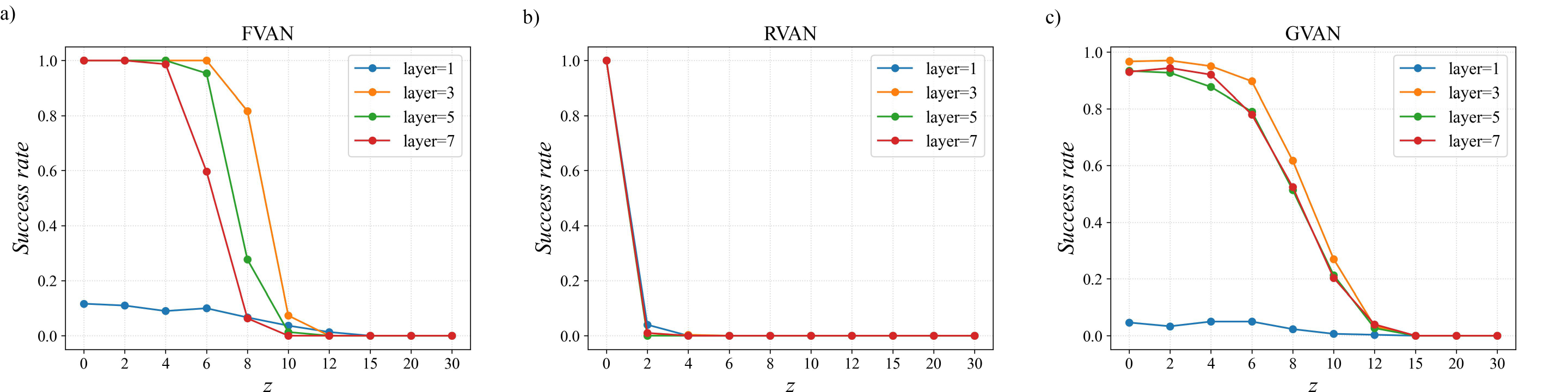}
\caption{\label{fig2} The success rates of finding the ground state vary with the number of layers on the WPE, with $N=60$ and $\alpha=0.2$. (a) The VAN framework based on FNN. (b) The VAN framework based on RNN. (c) The VAN framework based on GNN.}
\end{figure*}

\begin{figure*}
\includegraphics[scale=0.58]{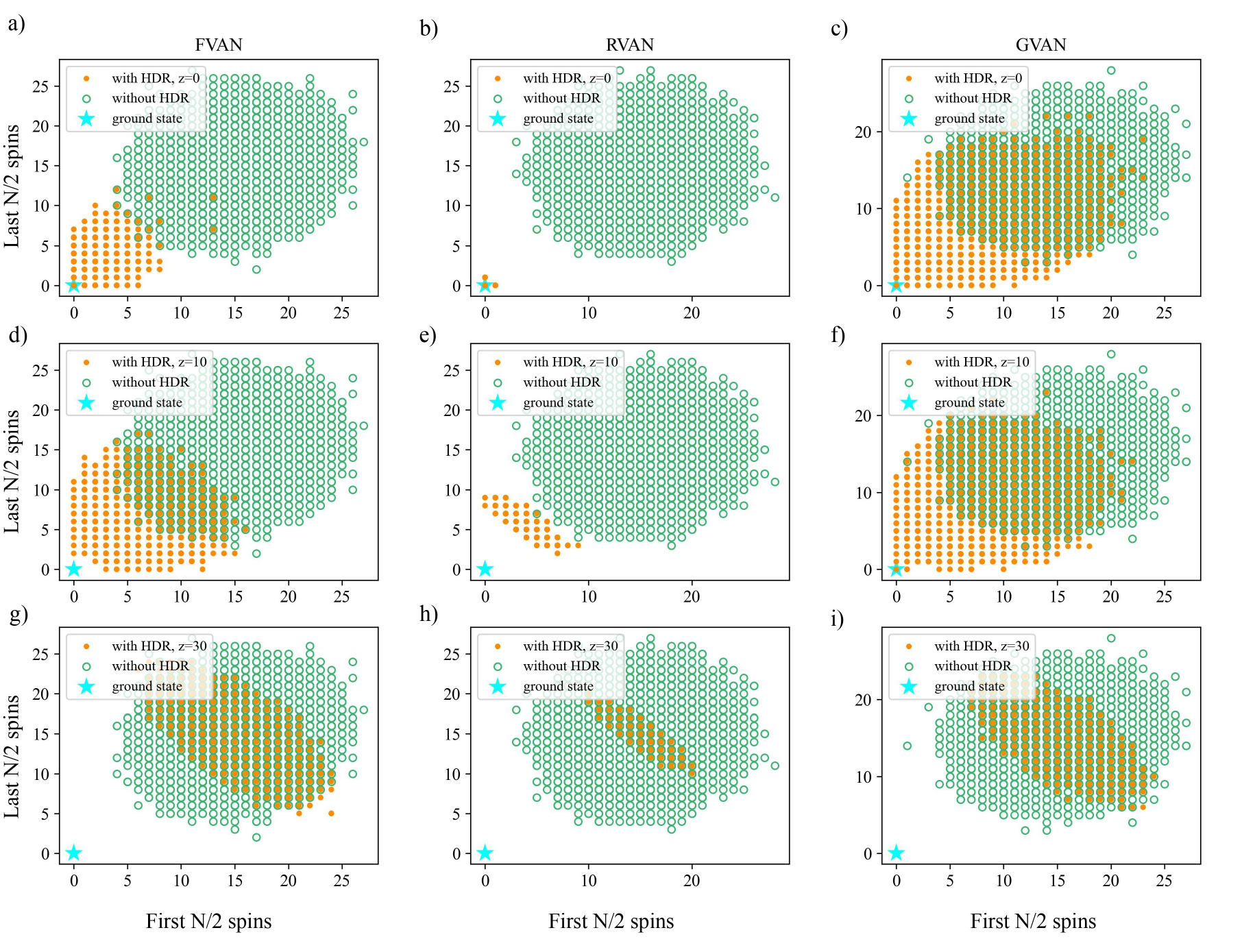}
\caption{\label{fig3} Scatter plot of the Hamming distance of the samples drawn after training and the ground state, with the X-axis and Y-axis denoting the Hamming distance of the first and last N/2 spins, respectively, when on the first WPE instance in Fig.~\ref{fig2} with system size $N=60$. (a) The VAN framework based on FNN, when with the Hamming distance regularizer (HDR) and $z=0$ and without the regularizer. (b) The VAN framework based on RNN, when with the Hamming distance regularizer and $z=0$ and without the regularizer. (c) The VAN framework based on GNN, when with the Hamming distance regularizer and $z=0$ and without the regularizer. (d)-(f) The VAN framework based on network architectures in (a)-(c), respectively, when with the Hamming distance regularizer and $z=10$ and without the regularizer. (g)-(i) The VAN framework based on network architectures in (a)-(c), respectively, when with the Hamming distance regularizer and $z=30$ and without the regularizer.}
\end{figure*}

\begin{figure*}
\includegraphics[scale=0.55]{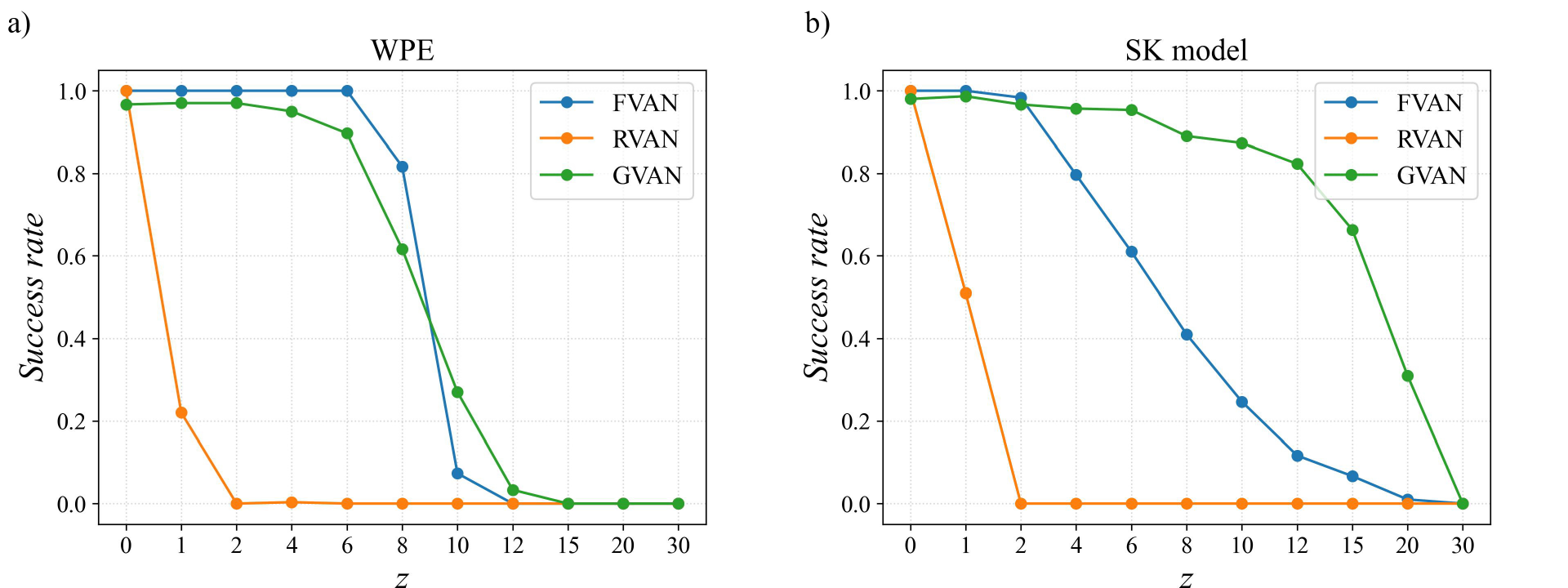}
\caption{\label{fig4} The success rates of finding the ground state vary with the value of $z$. (a) On the WPE, with system size $N=60$ and $\alpha=0.2$. (b) On the SK model, with system size $N=60$.}
\end{figure*}

We first conduct experiments to discuss the effect of the number of layers on the generalization capabilities of network architectures, including FNN, RNN, and GNN, combined with VAN. We emphasize that the combinations of network architectures and VAN frameworks may be varied, and their performance may be very different. Therefore, we select those combinations that seem to have the best performance based on the existing models in the literature as the methods we tested. Specifically, the combination of FNN and VAN is taken from Ref.~\cite{Wu2019VAN} and denoted as feed-forward variational autoregressive networks (FVAN) in this paper. The combination of RNN and VAN is taken from Ref.~\cite{Hibat2021vca} and denoted as recurrent variational autoregressive networks (RVAN), which is named variational classical annealing in Ref.~\cite{Hibat2021vca}. The combination of GNN and VAN is taken from Ref.~\cite{qunlong2023} and denoted as graph variational autoregressive networks (GVAN), which is named message passing variational autoregressive networks in Ref.~\cite{qunlong2023}.

The experiments are conducted on a planted model, Wishart planted ensemble (WPE) \cite{WPE2020}. The WPE is a class of fully connected Ising models with an adjustable hardness parameter $\alpha$ and planted solutions. The Hamiltonian of the WPE is defined as
\begin{equation}
    \label{eq5}
    H=-\frac{1}{2}\sum_{i\neq j}{J_{ij}s_{i}s_{j}},
\end{equation}
where the coupling matrix $\{J_{ij}\}$ is a symmetric matrix satisfying copula distribution. More details about WPE can be found in Ref.~\cite{WPE2020}.

The WPE instances we used here are hard to find the ground state without the regularizer, i.e., we can not distinguish the generalization capabilities of different network architectures when there is no regularizer in the loss function. Since there are no canonical datasets for the Ising models we are concerned with, we always generate 30 instances with the same model parameter and run the same method 10 times on each instance to exclude the effects of occasionality as much as possible and to reflect the general properties of network architectures.

The success rates of finding the ground state varying with the number of layers are shown in Fig.~\ref{fig2}. It can be seen that the number of layers has a critical impact on the success rates for the same value of $z$, which together compose the indicator to measure the generalization capabilities of generative models for solving intractable Ising models. For FNN and GNN under the VAN framework, the performance is best when the number of layers is three, and the success rates are higher than others when the value of $z$ is large. It indicates that FNN and GNN have the strongest generalization capabilities when the number of layers is three. The results for RNN under the VAN framework are almost the same, probably because it has strong expression capabilities to imitate the target distribution from the training datasets at any number of layers. In theory, deeper neural networks may perform better than shallow networks due to having more trainable parameters and dynamical eigenvalues \cite{Vanchurin2021}. However, the results in Fig.~\ref{fig2} indicate that the performance of the 3-layer networks is stronger than that of the 5-layer and 7-layer networks in terms of generalization capabilities. Considering the existing methods based on VAN \cite{Wu2019VAN, Hibat2021vca, Panfeng2021, qunlong2023}, we infer that the 3-layer networks already have a sufficient number of trainable parameters and dynamical eigenvalues to achieve good performance. Of course, it is not ruled out the possibility that 5-layer or 7-layer networks could achieve better performance than 3-layer networks, which may require more detailed hyperparameters adjustments to achieve.

We also find that the success rates of the VAN framework based on FNN and GNN decrease significantly as the value of $z$ increases to 8 and 10. The $z=12$ is the largest Hamming distance between the training datasets and the ground state that can be generalized by the above three methods when solving the WPE with system size $N = 60$ and $\alpha = 0.2$.

Next, we compare the Hamming distance between the ground state and the samples drawn from networks after training with and without the Hamming distance regularizer in the loss function. As shown in Fig.~\ref{fig3}, when with a large overlap between the ground state and the training datasets, i.e., a small value of $z$, the samples concentrate on the region close to the ground state as measured by Hamming distance. As the value of $z$ gets bigger, the sample distribution with the regularizer becomes similar to the sample distribution without the regularizer, where the ground state cannot be found. It can be found that the Hamming distance between the samples and the ground state is always concentrated close to the value of $z$, which demonstrates the strong constraints of the regularizer on the training datasets and final samples.

At the same time, the sample distribution with the regularizer and the same value of $z$ across network architectures can be significantly different. Based on sample distribution, we can have an intuitive qualitative comparison of the generalization capabilities of generative models in solving such intractable Ising models. It can be found that for the same $z$, the sample distribution of RVAN is always more concentrated, while the sample distribution of FVAN and GVAN is more dispersed. We infer that it is due to the strong capabilities of RNN to imitate the target distribution, and thus RVAN always quickly learns the features of the potential distribution of training datasets, and then mode collapse occurs quickly. In contrast, the mode collapse of FVAN and GVAN occurs later. Therefore, RVAN has the weakest ability to generate new samples that have not appeared in the training datasets, that is, its generalization capabilities are the weakest. It can also be inferred from Fig.~\ref{fig3}(a) and Fig.~\ref{fig3}(c) that GVAN has stronger generalization capabilities than FVAN, which has been validated in terms of the success rates of the three-layer networks in Fig.~\ref{fig2}. The results of samples sampled from the networks at more temperatures can be found in Supplementary Material~\ref{appen1}.

\begin{figure*}
\includegraphics[scale=0.55]{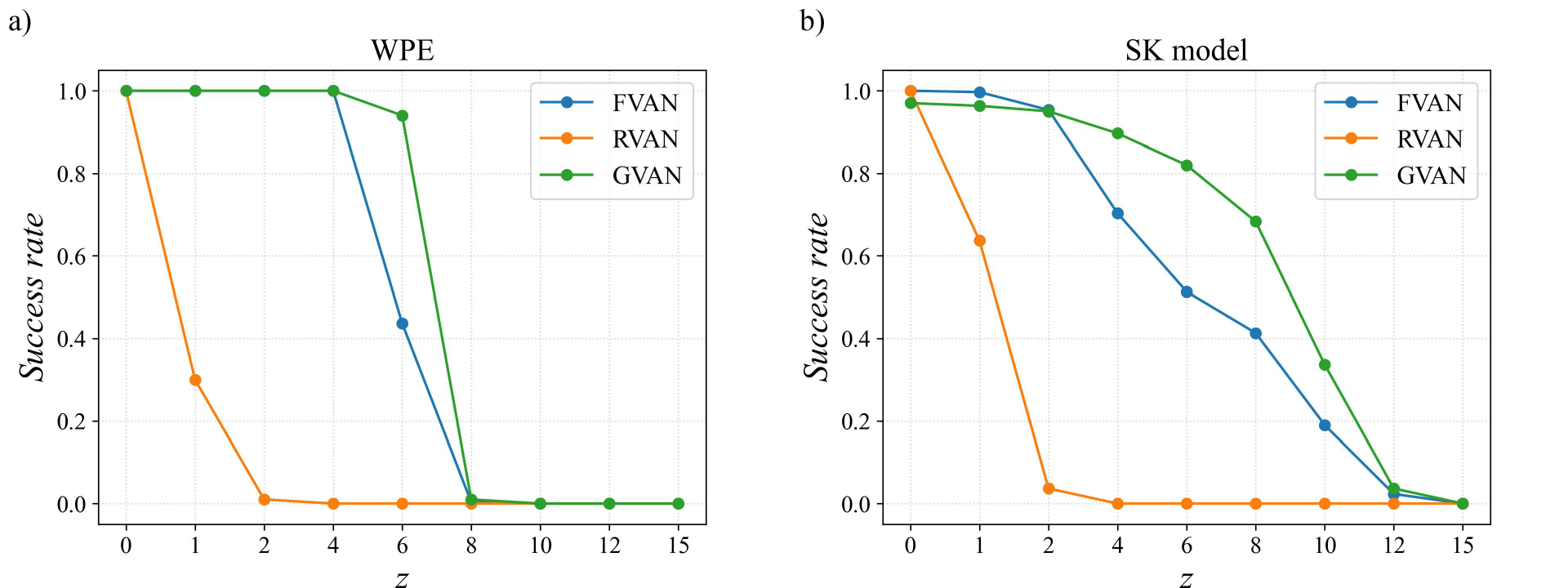}
\caption{\label{fig5} The success rates of finding the ground state vary with the value of $z$ on models with small system sizes. (a) On the WPE, with system size $N=30$ and $\alpha=0.1$. (b) On the SK model, with the system size $N=30$.}
\end{figure*}

In the following, we show the success rates varying with the value of $z$ from network architectures on the same models. As shown in Fig.~\ref{fig4}(a), the success rates of network architectures are significantly different on the WPE. The VAN framework based on GNN shows great potential, especially when the value of $z$ is large, which shows it may have the strongest generalization capabilities for the WPE. Also, we find that when the value of $z$ is small, such as $z=0$, the VAN framework based on FNN and RNN is better than with GNN, and it may be because FNN and RNN have stronger expression capabilities to imitate the target distribution from the training datasets than GNN.

We also experiment on the Sherrington-Kirkpatrick (SK) model \cite{Sherrington1975SK}, which is one of the most classical fully connected Ising spin glasses \cite{Panchenko2012TheSM, Panchenko2013}. Its Hamiltonian form is the same as Eq.~(\ref{eq5}), but $\{J_{ij}\}$ is from the Gaussian distribution with variance $1/N$ and is a symmetric matrix. We can exactly get the ground state by spin-glass Server (SGS) \cite{SGS} when system size $N\leq100$.

The results on the SK model are shown in Fig.~\ref{fig4}(b). Unlike the results on the WPE, the success rates are pretty high because it is easier to find the ground state of the SK model than the WPE with the same system size (when the WPE with $\alpha=0.2$). The difference in the success rates between the WPE and the SK model may be caused by their diverse distribution of the Hamming distance between excited states and the ground state \cite{Mehta2022, Valle2021}. Also, when the value of $z$ is large, the VAN framework based on GNN has the highest success rates, indicating that it has the strongest generalization capabilities.

Next, we are concerned with a question: can the quantification of the generalization capabilities of networks on small-scale problems be used to predict their performance on large-scale problems?

Therefore, we conduct experiments on the same models in Fig.~\ref{fig4} with small system sizes, and the results are shown in Fig.~\ref{fig5}. It can be found that for all two models, the relative performance of different network architectures on small-scale problems is almost identical to that of large-scale problems. These results indicate that the relative generalization capabilities of models with large system sizes can be predicted by the quantification of models with small system sizes. It is of great significance to search for optimal network architectures for large-scale problems.

Finally, we propose a quantitative metric that combines $z$ and success rates to quantify the generalization capabilities of network architectures. For the same $N$, it is defined as
\begin{equation}
    \label{eq6}
    Gen=\sum_{k=0}^{\lfloor N/2 \rfloor}{2^{k} SR_{k}},
\end{equation}
where $SR_{k}$ denotes the success rates when $z=k$. The Eq.~(\ref{eq6}) combines the size of the overlaps between the ground state and the training datasets and the corresponding success rates. Under this definition, the quantitative metrics for network architectures on the WPE and the SK model are shown in Tab.~\ref{table1}.

\begin{table}[h]
\caption{\label{table1} The quantitative metrics for network architectures on the WPE and the SK model corresponding to the results of Fig.~\ref{fig4} and Fig.~\ref{fig5}.}
\begin{ruledtabular}
\begin{tabular}{lccc}
 model & FVAN & RVAN & GVAN\\ \hline
 WPE, $N=60$ & 371.2 & 1.5  & \textbf{650.3} \\
 WPE, $N=30$ & 51.8 & 1.6 & \textbf{85.7} \\
 SK, $N=60$ & 12472.2 & 2.0  & \textbf{340506.0} \\
 SK, $N=30$ & 446.9 & 2.4 & \textbf{743.3} \\
\end{tabular}
\end{ruledtabular}
\end{table}

It can be seen that the relative sizes of the quantitative metrics of different network architectures on small-scale problems are almost the same as on large-scale problems. Therefore, we can use the generalization capabilities of network architectures on models with small system sizes to predict the relative capabilities of models with large system sizes.

We also calculate the quantitative metrics for network architectures with different layers in Tab.~\ref{table2}, where the $Gen$ gets the highest scores when the number of layers is three except RVAN.

\begin{table}[h]
\caption{\label{table2} The quantitative metrics for network architectures with different layers on the WPE corresponding to the results of Fig.~\ref{fig3}.}
\begin{ruledtabular}
\begin{tabular}{lccc}
 model & FVAN & RVAN & GVAN\\ \hline
layer=1 & 117.6 & \textbf{1.2}  & 30.6 \\
layer=3 & \textbf{369.1} & 1.1 & \textbf{648.3} \\
layer=5 & 166.5 & 1.0  & 528.3 \\
layer=7 & 75.2 & 1.0 & 575.3 \\
\end{tabular}
\end{ruledtabular}
\end{table}

\section{CONCLUSION AND DISCUSSIONS}
\label{sec4}

In summary, we design a Hamming distance regularizer based on variational autoregressive networks to quantify the generalization capabilities of network architectures. We also propose a quantitative metric that combines the size of the overlaps between the ground state and the training datasets generated by networks with the success rates of finding the ground state to achieve quantification. We conduct numerical experiments on several prototypical network architectures combined with VAN, such as FNN, RNN, and GNN, to quantify the generalization capabilities when solving intractable Ising models, including the WPE and the SK model. We also consider the effect of a hyperparameter, the number of layers of network architectures, on generalization capabilities.

Although we have only experimented on the FNN, RNN, and GNN in the framework of VAN, we must emphasize that any network architecture, when combined with VAN to solve intractable Ising models, can quantify the generalization capabilities using our method.

A recent study that has received a lot of attention is Neural Architecture Search \cite{Elsken2019, wistuba2019, mellor21a}. The results that the relative generalization capabilities of network architectures on large-scale problems can be predicted by their performance on small-scale problems demonstrate that our method can assist in searching for the optimal network architectures when solving large-scale problems, which may take a large amount of time in traditional search processes.

\begin{acknowledgments}
This work is supported by Project 61972413 (Z.M.) of National Natural Science Foundation of China.\\
\end{acknowledgments}

\noindent\textbf{Author contributions}:

M.G. conceived and designed the project. Z.M. and M.G. managed the project. Q.M. performed all the numerical calculations and analyzed the results. M.G., Q.M., and Z.M. interpreted the results. Q.M. and M.G. wrote the paper.\\

\noindent\textbf{Data availability}:

The data that support the findings of this study are available from the corresponding author upon reasonable request.\\

\noindent\textbf{Code availability}:

The code that supports the findings of this study is available from the corresponding author upon reasonable request.\\

\noindent\textbf{Competing interests}:

The authors declare no competing interests.\\

\bibliography{MLST-101874-HDR}

\newpage \onecolumngrid \newpage { \center \bf \large { Supplemental Material for: \\ A method for quantifying the generalization capabilities of generative models\\ for solving Ising models} \vspace*{0.1cm}\\  \vspace*{0.0cm} } 

\begin{center} Qunlong Ma,$^{1}$ Zhi Ma,$^{1}$ and Ming Gao$^{1}$ \\ 
\vspace*{0.15cm}
\small{\textit{$^{1}$ Henan Key Laboratory of Network Cryptography Technology, Zhengzhou 450001, China}} \\
\vspace*{0.25cm} 
\end{center}

\setcounter{section}{0}

\setcounter{equation}{0} 
\setcounter{figure}{0} 
\setcounter{page}{1} 
\makeatletter 
\renewcommand{\theequation}{S\arabic{equation}} 
\renewcommand{\thefigure}{S\arabic{figure}} 
\renewcommand{\bibnumfmt}[1]{[S#1]} 
\renewcommand{\citenumfont}[1]{#1} 

\section{More results sampled at different temperatures}
\label{appen1}

In this section, we show the scatter plot of the Hamming distance of the samples drawn after training and the ground state, similar to Fig.~\ref{fig3} of the main text but at more temperatures.

\begin{figure}[h]
\includegraphics[scale=0.44]{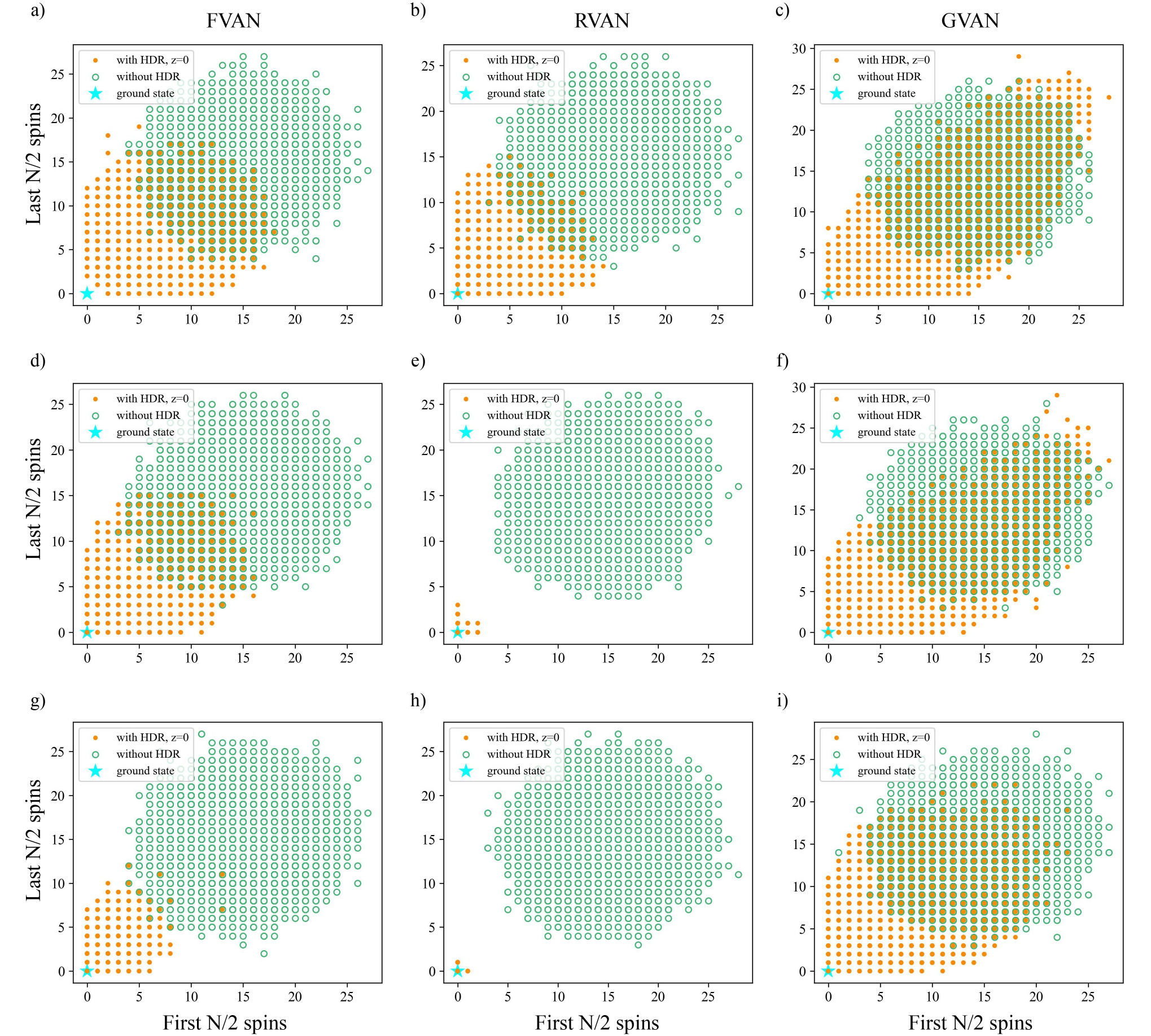}
\caption{\label{figS1} Scatter plot of the Hamming distance of the samples drawn after training and the ground state, with the X-axis and Y-axis denoting the Hamming distance of the first and last N/2 spins, respectively, when on the same WPE instance as Fig.~\ref{fig3} in the main text with system size $N = 60$. (a) The VAN framework based on FNN, when with the Hamming distance regularizer (HDR) and $z = 0$ and without the regularizer at inverse temperature $\beta=1.5$. (b) The VAN framework based on RNN, when with the Hamming distance regularizer and $z = 0$ and without the regularizer at inverse temperature $\beta=1.5$. (c) The VAN framework based on GNN, when with the Hamming distance regularizer and $z = 0$ and without the regularizer at inverse temperature $\beta=1.5$. (d)-(f) The VAN framework based on network architectures and HDR in (a)-(c), respectively, when at inverse temperature $\beta=5.5$. (g)-(i) The VAN framework based on network architectures and HDR in (a)-(c), respectively, when at inverse temperature $\beta=20.5$. The results of Fig.~\ref{fig3} of the main text are with inverse temperature $\beta=20.5$.}
\end{figure}

\begin{figure}[h]
\centering
\includegraphics[scale=0.6]{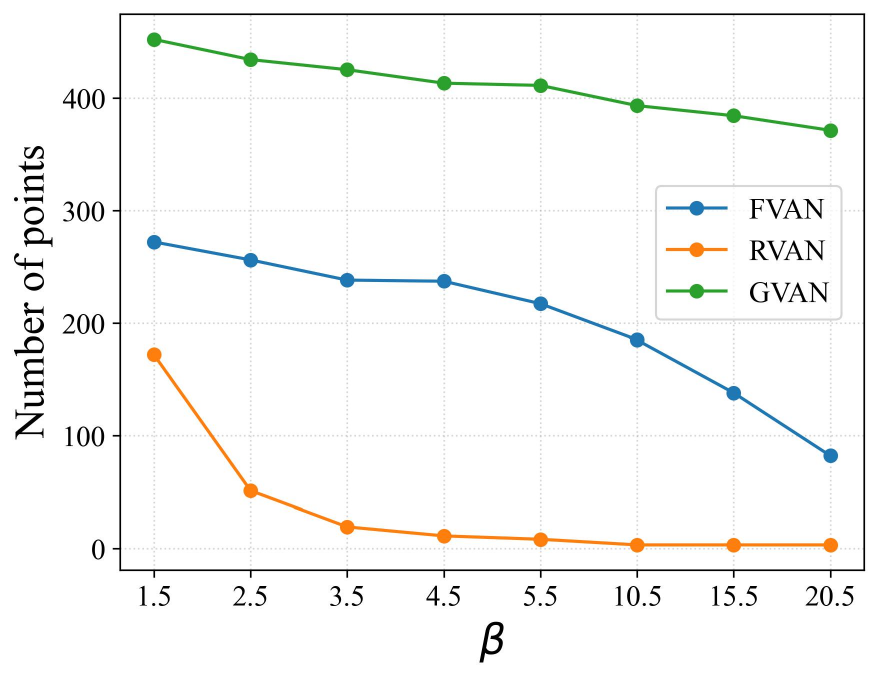}
\caption{\label{figS2} The number of different points contained in ten thousand samples under multiple inverse temperatures, with HDR and $z=0$, when on the same WPE instance as Fig.~\ref{fig3} of the main text with system size $N = 60$.}
\end{figure}

At the same time, we also count the number of different points in the scatter plots similar to Fig.~\ref{fig3} and Fig.~\ref{figS1} at more inverse temperatures, as shown in Fig.~\ref{figS2}.

According to Fig.~\ref{fig3}, Fig.~\ref{figS1} and Fig.~\ref{figS2}, it can be found that as the temperature decreases, the distribution of all three networks gradually concentrates around the configuration with Hamming distance of $z$ from the ground state. It is because that including HDR in the loss function can control the training datasets to be distributed as much as possible on that configuration, and we name it the target state. Among them, RVAN has the fastest speed of mode collapse, which means it can learn the target distribution the fastest.

Due to the rapid mode collapse of RVAN, its ability to generate samples other than the target state is weak. In contrast, the mode collapse of FVAN and GVAN occurs later, and thus RVAN has the weakest ability to generate new samples that have not appeared in the training datasets, that is, its generalization capabilities. This is particularly evident as $z$ gradually increases. For example, in Fig.~\ref{fig4} of the main text, when $z=2$, the success rate of RVAN in finding the ground state approaches 0. At the same time, the average Hamming distance between the target state (or, training datasets) and the ground state is only 2, which means that only 2 spins need to be flipped to obtain the ground state. In addition, from the collective behavior of FVAN and GVAN samples with HDR ($z = 0$) in Fig.~\ref{fig3} of the main text, we infer that they have stronger generalization capabilities than RVAN, which is also validated in terms of quantitative metrics in Tab.~\ref{table1} of the main text.

\end{document}